\documentstyle [12pt]{article}
\title{
\hspace{3.0truein}{\small IFT-478-UNC}\\
\vspace{-0.1truein}
\hspace{3.0truein}{\small NUP-A-93-15}\\
\vspace{0.2truein}
{Equivalence of Several\\ Chern-Simons Matter Models}
}
\author{Wei Chen\footnotemark[1]\\
Department of Physics\\
University of North Carolina\\
Chapel Hill, NC 27599-3255\\
and\\
          Department of Physics\\
          University of British Columbia\\
          Vancouver, B.C. Canada V6T 1Z1\\
\\
           Chigak Itoi\footnotemark[2]\\
               Department of Physics and Atomic Energy Research Institute\\
College of Science and Technology, Nihon University\\
Kanda Surugadai, Chiyoda-ku, Tokyo 101, Japan}
\date{PACS numbers: 11.15.Bt, 11.15.Pg}
\footnotetext{$^*$ chen@physics.unc.edu}
\footnotetext{$^\dagger$  itoi@phys.cst.nihon-u.ac.jp}

\begin{document}
\maketitle
\vspace{0.2truein}
\begin{abstract}
Not only does Chern-Simons (CS) coupling
characterize statistics, but
also spin and scaling dimension of matter fields.
We demonstrate
spin transmutation in relativistic
CS matter theory, and moreover
show equivalence of several models. We study CS vector
model in some details, which provide consistent check
to the assertion of the equivalence.
\end{abstract}
\newpage
\baselineskip=22.0truept

Introducing CS interaction in two spatial dimensions
is equivalent to attaching to particles 'magnetic' flux tubes.
Considerable interests focus recently on strong
CS couplings so that each electron
carries, for instance, two flux tubes or more.
This serves as a key ingredient
in a recent successful theory
of the half-filled Landau level to
understand the quantum Hall effects with the filling factors
$\nu=1/2$, and $ = p/(2p+1)$ (p an integer) \cite{HLR,KZ}.
However, as the interaction is getting strong,
perturbation theory which has being used in many cases
can not be reliable, in general.
In this Letter, we suggest alternatives of the model with Dirac
fermion coupled to CS, in the relativistic formalism.
We shall show
the following three $U(1)$ quantum gauge field theory models,
which look apparently so different,
are equivalent one to the others. These are
\begin{equation}
{\cal L}_F =  \bar\psi
[-i\sigma_\mu(\partial_\mu + ia_\mu +iC_\mu) +M]\psi
- \frac{i}{8\pi\alpha}
\epsilon_{\mu\nu\lambda}a_\mu\partial_\nu a_\lambda\;,
\label{SF}
\end{equation}
where $\psi$ is a two-component Dirac field,
$\sigma_{\mu}$ ($\mu=1,2,3$) Pauli matrices,
and $C_\mu$ an external gauge field;
\begin{equation}
{\cal L}_B =
\frac{1}{2}B^*_{\mu}
[-i\epsilon_{\mu\nu\lambda}(\partial_\nu + ia_\nu+iC_\nu)
+M\delta_{\mu\lambda}]B_\lambda
- \frac{i}{8\pi(\alpha-\frac{1}{2})}
\epsilon_{\mu\nu\lambda}a_\mu\partial_\nu a_\lambda\;,
\label{SB}
\end{equation}
where $B_\mu$ is a complex vector field; and
\begin{equation}
{\cal L}_{FF} =  \bar\Psi
[-i\frac{2}{3}L_\mu(\partial_\mu + ia_\mu +iC_\mu) +M]\Psi
- \frac{i}{8\pi(\alpha-1)}
\epsilon_{\mu\nu\lambda}a_\mu\partial_\nu a_\lambda\;,
\label{SFF}
\end{equation}
where $\Psi$ is a spin-$(3/2)$ field,
and $L_\mu$ $4\times4$ matrices.
While it is of the most interest to
explore intrinsic relations among different models, obviously
(\ref{SF}), (\ref{SB}), and (\ref{SFF}) can be good
perturbation theories only around
$\alpha = 0, 1/2,$ and $1$, respectively.

Our expectation of the equivalence is based on the observation
that the CS coefficient characterizes not only the
statistics, but also the spins of the coupled matter fields.
Therefore, it is possible to trade
CS coupling for higher spins, and vice versa.
To understand the mechanism of spin transmutation precisely,
we calculate the partition functions of the three models,
by using the particle path integral method \cite{I,SSS}.
We obtain, for any given $\alpha$,
\begin{equation}
Z_F[C] = Z_B[C] = Z_{FF}[C]\;.
\label{EQ}
\end{equation}

The CS coupling characterizes as well the
scaling dimensions of operators
and other physical quantities \cite{CSW,CL,WW,CFW}.
The equivalence seen here
allows perturbative expansion of the quantum CS
field theory in one of its versions
with weak CS coupling.
In the second half of the Letter,
we examine the model (\ref{SB}) in some
details, while leave further investigation of
(\ref{SFF}) to
a separate publication \cite{CI1} except a comment:
the model (\ref{SFF}) around $\alpha = 1$
might provide a proper description of
the half-filled Landau level,
as it contains fermion particles carrying two
flux tubes with small perturbation in $\alpha -1$.
Some aspects of the model (\ref{SB}) in a slightly
different form
(where the $B_\mu$ field was real)
were discussed
in the literature \cite{DJ,H,HPK}. 
Here we look into some others, which
provide a consistent check to the assertion of the
equivalence.
Moreover, combining with the known results
about the model (\ref{SF}) \cite{CSW},
we shall discuss the scaling dimension of the matter field
against the statistical parameter $\alpha$ in the regin $[0,1/2]$.

Deriving (\ref{EQ}), we start from the partition function
$Z_F[C]=\int{\cal D}a_\mu
           {\cal D}\bar{\psi}
           {\cal D}\psi e^{-S_F}.$
The path integral over fermion fields is of Gaussian type,
so it is readily to obtain
\begin{eqnarray}
Z_F[C]&=&\int{\cal D}a_\mu\sum^\infty_{n=0}
\frac{1}{n!}(-W_F)^n
 exp[\frac{i}{8\pi\alpha}\int d^3x
\epsilon^{\mu\nu\lambda}a_\mu\partial_\nu a_\lambda]\;,\\
W_F&=&-Trlog[-i(\partial_\mu+ia_\mu+iC_\mu)\sigma_\mu+M]\;.
\end{eqnarray}
To integrate over $a_\mu$,
we use a particle path integral representation of $W_F$ \cite{I,SSS},
\begin{eqnarray}
W_F &=& \int{\cal D}{\bf X}
exp\{-\int^1_0 dt[M\sqrt{\dot{{\bf X}}^2}
+i{\bf a}\cdot\dot{\bf X}
+i{\bf C}\cdot\dot{\bf X}]
+\frac{i}{2}\Phi[ \frac{\dot{{\bf X}}}{|\dot{{\bf X}}|}]\},
\label{WF}\\
 \Phi[{\bf e}] &=& \int_D du ds{\bf e}\cdot
[\partial_s{\bf e}\times\partial_u{\bf e}]\;.
\label{phi}
\end{eqnarray}
(\ref{phi}) is defined as spin factor. Geometrically, this
is the area enclosed by a path ${\bf e}(t)$ ($0\leq t \leq 1$)
on a unit sphere $S^2$. The numerical coefficient
of the spin factor in (\ref{WF}) reflects the spin
of the matter, which is apparently $1/2$ for the Dirac field.
Now the integral over $a_\mu$ is readily to perform, and it gives
\begin{equation}
Z_F[C]=\sum^\infty_{n=1}\frac{(-1)^n}{n!}\int\prod^n_{i=1}
  {\cal D}{\bf X}_i
exp\{-\sum^{n}_{i=1}[\int dt (M\sqrt{\dot{\bf X}_i^2}
+i{\bf C}\cdot\dot{\bf X}_i)
-\frac{i}{2}\Phi[\frac{\dot{\bf X}_i}{|\dot{\bf X}_i|}]
-i\frac{\alpha}{2}\Theta_{ii}]
+i\alpha\sum_{i<j}\Theta_{ij}\}\;,
\label{SFa}
\end{equation}
\begin{eqnarray}
& &\Theta_{ij} =\frac{1}{\alpha}
\int^1_0 dt\int^1_0 ds\frac{dX_i^\mu}{dt}\frac{dX^\nu_j}{ds}
<a_\mu(X_i)a_\nu(X_j)>\;,\\
& & <a_\mu(x)a_\nu(y)>
= 8\pi\alpha\epsilon_{\mu\nu\lambda}\frac{x^\lambda-y^\lambda}{|{\bf x}-
{\bf y}|^3}\;.\label{CS}
\end{eqnarray}
(\ref{CS}) is the CS propagator in the Landau
gauge.

Now, we come to the key point for spin transmutation.
It is shown in \cite{I,SSS} that
 the self-energy $\Theta_{ii}$ is related to the spin factor $\Phi$
(see (\ref{phi})) via
\begin{equation}
\Theta_{ii}-2\Phi = 4\pi ~~~(mod~~ 8\pi)\;;
\label{psiii}
\end{equation}
and that the relative energy
$\Theta_{ij}$ $(i\neq j)$ is the Gaussian linking number
\begin{equation}
\Theta_{ij} \in 4\pi{\bf Z}\;.
\label{psiij}
\end{equation}
Using (\ref{psiii}) and (\ref{psiij}), we shift
$\alpha$ in
(\ref{SFa}) by $ - 1/2$, and then $Z_F[C]$ takes a form
\begin{equation}
Z_F[C]=\sum^\infty_{n=1}\frac{1}{n!}\int\prod^n_{i=1}
  {\cal D}{\bf X}_i
exp\{-\sum^{n}_{i=1}[\int dt( M\sqrt{\dot{\bf X}_i^2}
+i{\bf C}\cdot\dot{\bf X}_i)
-i\Phi[\frac{\dot{\bf X}_i}{|\dot{\bf X}_i|}]
-i\frac{\alpha-\frac{1}{2}}{2}\Theta_{ii}]
+i(\alpha-\frac{1}{2})\sum_{i<j}\Theta_{ij}\}\;.
\label{ZFb}
\end{equation}
Notice the numerical coefficient of the spin factor $\Phi$
is now $1$ in (\ref{ZFb}), replacing $1/2$ in (\ref{SFa}).
Recovered the integral
over $a_\mu$, (\ref{ZFb}) is written as
\begin{eqnarray}
Z_F[C]&=&\int{\cal D}a_\mu\sum^\infty_{n=0}
\frac{1}{n!}(-W_B)^n
 exp[\frac{i}{8\pi(\alpha-1/2)}\int d^3x
\epsilon_{\mu\nu\lambda}a_\mu\partial_\nu a_\lambda]\;,
\label{ZFc}\\
W_B&=& \int{\cal D}{\bf X}
exp\{-\int^1_0 dt[M\sqrt{\dot{{\bf X}}^2}
+i{\bf a}\cdot\dot{\bf X}
+i{\bf C}\cdot\dot{\bf X}]
+i\Phi[ \frac{\dot{{\bf X}}}{|\dot{{\bf X}}|}]\},
\label{WB}\\
&=&-Trlog[-(\partial_\mu+ia_\mu+iC_\mu)L_\mu+M]\;,
\end{eqnarray}
where $L^\mu_{\nu\lambda}$, chosen here
as $-i\epsilon_{\mu\nu\lambda}$, is the
spin matrix with the eigenvalue $1$. The right hand side of
(\ref{ZFc}) is nothing but the partition function of (\ref{SB}).
To continue our derivation of (\ref{EQ}),
we use (\ref{psiii}) and (\ref{psiij}) once again, and shift
$(\alpha-1/2)$ in (\ref{ZFb}) by $-1/2$ to $(\alpha-1)$.
We obtain
\begin{eqnarray}
Z_F[C]&=&\int{\cal D}a_\mu\sum^\infty_{n=0}
\frac{1}{n!}(-W_{FF})^n
 exp[\frac{i}{8\pi(\alpha-1)}\int d^3x
\epsilon^{\mu\nu\lambda}a_\mu\partial_\nu a_\lambda]\;,\label{SFFa}\\
W_{FF}&=& \int{\cal D}{\bf X}
exp\{-\int^1_0 dt[M\sqrt{\dot{{\bf X}}^2}
+i{\bf a}\cdot\dot{\bf X}
+i{\bf C}\cdot\dot{\bf X}]
+i\frac{3}{2}\Phi[ \frac{\dot{{\bf X}}}{|\dot{{\bf X}}|}]\},
\label{WFF}\\
&=&-Trlog[-(\partial_\mu+ia_\mu+iC_\mu)L_\mu+M]\;,
\end{eqnarray}
where the eigenvalue of
${\bf L}^2$ is
$(3/2)(3/2+1)$. The right hand side of (\ref{SFFa}) is
just the partition function of (\ref{SFF}). In summary,
Lagrangians (\ref{SF}), (\ref{SB}), and (\ref{SFF})
are just different
versions of the same CS matter quantum field theory.
It is not difficult to repeat the above procedure for more
versions with higher spin matters.
As a special case, when $\alpha$ is an integer or half integer
and  $C_\mu$ absent,
the CS matter theory turns out to be
a free theory of spin $|1/2+\alpha|$ particles.

Next, we discuss the model (\ref{SB})
(the external field $C_\mu$ is ignored hereafter).
Like (\ref{SF}), (\ref{SB})
is invariant under $U(1)$ gauge transformations:
$a_\mu \rightarrow a_\mu - \partial_\mu\Lambda,
{}~ B_\mu \rightarrow e^{i\Lambda}B_\mu.$
The equations of motion which follow from (\ref{SB}) are
\begin{eqnarray}
& &\epsilon_{\mu\nu\lambda}\partial_\nu a_\lambda =4\pi(\alpha-1)j_\mu\;,
\label{eq2}\\
& &\epsilon_{\mu\nu\lambda}(\partial_\nu+ia_\nu)B_\lambda+iMB_\mu=0\;,
\label{eq1}
\end{eqnarray}
where the current $j_\mu=-i\epsilon_{\mu\nu\lambda}B^*_\nu B_\lambda$
(the equation for $B^*_\mu$,
similar to (\ref{eq1}), is omited).
{}From  (\ref{eq2}) and (\ref{eq1}), one readily obtains
the current conservation, $\partial_\mu j_\mu =0,$
and the constraint for $B_\mu$ field:
$(\partial_\mu+ia_\mu)B_\mu=0$.
Since $a_\mu$ has no independent degree of freedom (see (\ref{eq2})),
this constraint reduces the number of independent
degrees of freedom of $B_\mu$ field to $2$,
the same with that of the Dirac field in three dimensions.

The $B_\mu$ field propagator and
its inverse propagator are readily to obtain
\cite{I2}
\begin{equation}
D_{\mu\nu}=\frac{1}{p^2+M^2}(\epsilon_{\mu\nu\lambda}p_\lambda
+\frac{p_\mu p_\nu}{M}+M\delta_{\mu\nu})\;,
{}~~~~D^{-1}_{\mu\nu} = -\epsilon_{\mu\nu\lambda}p_\lambda +
M\delta_{\mu\nu}\;.
\end{equation}
The one-loop correction from the vector field $B_\mu$
to the polarization tensor of CS
$\Pi_{\mu\nu}=\Pi_e(p,M)(\delta_{\mu\nu}p^2-p_\mu p_\nu)
+\Pi_o(p,M)\epsilon_{\mu\nu\lambda}p_\lambda$
is finite
(by the regularization by dimensional reduction \cite{CSW}):
\begin {eqnarray}
\Pi_e(p,M) &=& \frac{1}{12\pi}\frac{1}{\sqrt{M^2}}
[1+{\cal O}(\frac{p^2}{M^2})]\;,\\
\Pi_o(p,M) &=& -\frac{1}{2\pi}sign(M)
[1+{\cal O}(\frac{p^2}{M^2})]\;.
\end{eqnarray}
This implies that, in the low energy 
limit,
the CS gauge field behaves like a dynamical,
topologically massive gauge field. This is exactly
what happens in the model of CS
coupled to  Dirac field, (\ref{SF}).
It is argued that, at two-loop and beyond,
the CS mass receives no further corrections, -- a natural
extention of the no-renormalization theorem \cite{CH,SSW}
to the CS vector model \cite{HPK}.

Like in the CS Dirac field model, at two-loop
the matter field, now the vector, self-energy
has a simple pole, which reflects the appearance of
log divergence at this order,
and therefore the $B_\mu$ field needs nontrivial
wave-function renormalization, from which
we obtain the scaling dimension of the $B_\mu$ field
\begin{equation}
[B_\mu] = 1 - \frac{(\alpha-1/2)^2}{8} + {\cal O}((\alpha-1/2)^4)\;.
\label{db}
\end{equation}

Since (\ref{SF}) and (\ref{SB}) are just different versions of
the same theory,
$\psi$ and $B_\mu$ describe the same matter
field for any given $\alpha$,
let's recall the scaling dimension of the Dirac field in model
(\ref{SF}), calculated in \cite{CSW},
$[\psi] = 1 - \frac{\alpha^2}{3} + {\cal O}(\alpha^4).$
On the other hand, it is natural to assume that
the scaling dimension of the matter field is a continuous function
of $\alpha$ in the region [$0$, $1/2$], for instance. Then,
since quantum fluctuation decreases the dimension of matter field
as $\alpha$ varies away from both the fermion
 ($\alpha = 0$) and boson ($\alpha = 1/2$) points,
there must exist at least one local minimum in
the scaling dimension of matter versus $\alpha$
between $\alpha = 0$ and $1/2$.

It is  interesting to observe that when $M=0$,
the $U(1)$ CS vector model (\ref{SB}) turns out to be
a topological non-Abelian $SU(2)$  gauge theory,
which has no dynamical degree of freedom!
To see this, we set
$ a_\mu=ea^1_\mu$, with $e=\sqrt{4\pi(\alpha-1/2)}$,
and $B_\mu = a_\mu^2+ia_\mu^3,$
substitute them into (\ref{SB}), then have
\begin{equation}
{\cal L} = -\frac{i}{2}\epsilon_{\mu\nu\lambda}
(a^a_\mu\partial_\nu a^a_\lambda
+\frac{e}{3}\epsilon^{abc}a^a_\mu a^b_\nu a^c_\lambda)\;.
\end{equation}
But, the equivalence seen above cannot be simply
extended to the massless case,
as applying the particle path integral method
to massless fields is somehow ambiguous.

To conclude the Letter, we remark the equivalence, however,
can be generalized
to the matter self-interactions in CS models.
For instance, adding a quartic interaction
$(\bar{\psi}\psi)^2$ in (\ref{SF}) can be seen \cite{CI1}
equivalent to adding a quartic interaction
$(B^*_\mu B_\mu)^2$
in the vector version of the CS theory, (\ref{SB}).

The authors thank I. Affleck, J.-W. Gan, G. Semenoff, and Y.-S. Wu
for discussions. The work of W.C. was supported in part by
the  U.S. DOE grant No. DE-FG05-85ER-40219,
and by NSERC, Canada. One of us (C.I.)
is grateful to G. Semenoff and University of British
Columbia for the kind hospitality extended to him.

\end{document}